\documentclass{article}

\usepackage[preprint,nonatbib]{neurips_2025}

\usepackage[utf8]{inputenc} 
\usepackage[T1]{fontenc}    
\usepackage{hyperref}       
\usepackage{url}            
\usepackage{booktabs}       
\usepackage{amsfonts}       
\usepackage{nicefrac}       
\usepackage{microtype}      
\usepackage{xcolor}         
\usepackage{stmaryrd} \SetSymbolFont{stmry}{bold}{U}{stmry}{m}{n}
\usepackage{amsmath,bm}
\usepackage[normalem]{ulem} 
\usepackage{graphicx}
\usepackage{centernot}
\usepackage{algorithm}
\usepackage{algpseudocode}
\usepackage{verbatim}
\usepackage{wrapfig}
\usepackage{booktabs}

\newcommand{\longoverbrace}[2]{\overbrace{#1}^{\text{\hbox to 0cm{\hss #2 \hss}}}}
\newcommand{\longunderbrace}[2]{\underbrace{#1}_{\text{\hbox to 0cm{\hss #2 \hss}}}}

\DeclareMathOperator{\trace}{trace}
\newcommand{\Tr}[1]{\trace\left(#1\right)}
\newcommand{\calP}{\mathcal{P}_T}
\newcommand{\opcalP}[1]{\calP\left(#1\right)}
\newcommand{\calA}{\mathcal{A}}
\newcommand{\calL}{\mathcal{L}}

\newcommand{\dotprod}[2]{\langle #1,#2 \rangle}
\newcommand{\EE}[1]{\mathbf{E}\left[#1 \right]}
\newcommand{\maxset}[1]{\max\left\{#1 \right\}}
\newcommand{\infset}[1]{\inf\left\{#1 \right\}}

\usepackage{amsfonts,amsmath,amssymb,amsthm,commath,bm}

\newtheorem{lemma}{Lemma}
\newtheorem{theorem}{Theorem}

\usepackage{stmaryrd}
\usepackage{trimclip}
\makeatletter
\DeclareRobustCommand{\shortto}{%
        \mathrel{\mathpalette\short@to\relax}%
}
\newcommand{\short@to}[2]{%
        \mkern2mu
        \clipbox{{.3\width} 0 0 0}{$\m@th#1\vphantom{+}{\shortrightarrow}$}%
}
\makeatother

\allowdisplaybreaks

\usepackage{xr-hyper}
\usepackage{newfloat}

\DeclareFloatingEnvironment[ 
        fileext=los,
        listname=List of Supplementary Figures,
        name=Supplementary Figure,
        placement=tbhp,
        within=none,
]{supfigure}

\DeclareFloatingEnvironment[ 
        fileext=los,
        listname=List of Supplementary Tables,
        name=Supplementary Table,
        placement=tbhp,
        within=none,
]{suptable}

\title{Fast Two-photon Microscopy by Neuroimaging with Oblong Random Acquisition (NORA)}

\author{
        Esther~Whang\\
        Department of Biomedical Engineering\\
        Johns Hopkins University\\
        Baltimore, MD 21218 \\
        \texttt{@jh.edu} \\
        \And
        Skyler Thomas \\
        Department of Biomedical Engineering\\
        Johns Hopkins University\\
        Baltimore, MD 21218 \\
        \texttt{jiyi@jhu.edu} \\
        \And
        Ji Yi \\
        Department of Biomedical Engineering\\
        Kavli Neurodiscovery Institute\\
        Johns Hopkins University\\
        Baltimore, MD 21218 \\
        \texttt{jiyi@jhu.edu} \\
        \And
        Adam S.~Charles\\
        Department of Biomedical Engineering\\
        Center for Imaging Science\\
        Kavli Neurodiscovery Institute\\
        Johns Hopkins University\\
        Baltimore, MD 21218 \\
        \texttt{adamsc@jhu.edu} \\
}

\begin{document}

\maketitle

\begin{abstract}
Advances in neural imaging have enabled neuroscientists to study how the joint activity of large neural populations conspire to produce perception, behavior and cognition. Despite many advances in optical methods, there exists a fundamental tradeoff between imaging speed, field of view, and resolution that limits the scope of neural imaging, especially for the raster-scanning multi-photon imaging needed for imaging deeper into the brain. One approach to overcoming this trade-off is computational imaging: the co-development of optics and algorithms where the optics are designed to encode the target images into fewer measurements that are faster to acquire, and the algorithms compensates by inverting the optical coding to recover a larger or higher resolution image. We present here one such approach for raster-scanning two-photon imaging: Neuroimaging with Oblong Random Acquisition (NORA). NORA quickly acquires each frame in a microscopy video by subsampling only a fraction of the fast scanning lines, ignoring large portions of each frame. NORA mitigates the loss of information by 1) extending the point-spread function in the slow-scan direction to effectively integrate the fluorescence of several lines into a single set of measurements and 2) imaging different, randomly selected, lines at each frame. Rather than reconstruct the video frame-by-frame, NORA recovers full video sequences via nuclear-norm minimization on the pixels-by-time matrix, for which we prove theoretical guarantees on recovery. We simulated NORA imaging using the Neural Anatomy and Optical Microscopy (NAOMi) biophysical simulator, and used the simulations to demonstrate that NORA can accurately recover 400~$\mu$m~X~400~$\mu$m fields of view at subsampling rates up to 20X, despite realistic noise and motion conditions. As NORA requires minimal changes to current microscopy systems, our results indicate that NORA can provide a promising avenue towards fast imaging of neural circuits.
\end{abstract}

\section{Introduction}

Discovering principles of neural computation at the cellular level rests on the ability to record large populations of single cell's activity. Two-photon microscopy enables the imaging of activity at a large scale deep within the tissue while maintaining high spatial resolution, resulting in its popularity within neuroscience. For standard multi-photon microscopes, the effective speed of such standard systems enable scanning of reasonably large field of views (FOV) on the order of 0.5-1~mm square FOVs at approximately 30~Hz. However, while $\approx$30Hz is sufficient for capturing dynamics from calcium indicators, the framerate is currently insufficient for capturing fast neural dynamics such as from voltage indicators, which requires at minimum ~x10 increase in frame rate to effectively record. The restriction on the frame rate arises due to how two-photon (2P) microscopes acquire images, a process called raster-scanning in which the FOV is scanned point-by-point. As a result, this 2P microscopy scanning method faces a trade-off between temporal and spatial resolution. Specifically, improving the frame rate requires reducing the amount of samples acquired per frame resulting in poorer spatial resolution, but fully sampling a FOV lowers the frame rate. Therefore, imaging a larger area imposes further restrictions on the temporal and spatial resolution.

Addressing this speed-resolution-area tradeoff is incredibly important to consider when increasing the size of our window into neural dynamics and has thus prompted numerous approaches for high-speed, large-scale imaging ~\cite{wu_speed_2021}. Most approaches develop purely optical designs that spread light across the tissue~\cite{lu2017video,song2017volumetric,demas2021high}. Examples include volumetric imaging via bessel beams~\cite{lu2017video} or stereoscopic imaging~\cite{song2017volumetric}, multiplane imaging through various forms of beam-splitting~\cite{demas2021high}, and ROI-specific sampling~\cite{mattison2023high}. Computational imaging methods, which utilize the co-design of hardware and software, have become another way to address the tradeoff between frame rate and spatial resolution ~\cite{kazemipour2019kilohertz}. While effective, these methods all require collecting light from each fluorescing object in the FOV. Given that the power constraints on \textit{in vivo} imaging systems and the scattering nature of the tissue, spreading the light too much will quickly reduce signal-to-noise levels beyond recoverability and thus effectively cap the maximum effective improvement of such approaches. Furthermore, many of these methods are vulnerable to motion artifacts due to the need for pre-established structural priors, restricting the generalizability of such systems.

We present here an ultra-fast 2P microscope that overcomes the speed-resolution-area tradeoff. Our method, Neuroimaging by Oblong Random Acquisition (NORA), only requires simple hardware and well-studied optimization-based reconstruction methods. It utilizes randomized line-by-line subsampling and an elongated point spread function (PSF) to efficiently encode measurements at each frame. In particular, light is not assumed to be collected from the full FOV, enabling faster subsampling by simply skipping portions of each frame. From the subsampled-and-blurry measurements, the full recording can then be reconstructed by posing the recovery as a matrix completion problem~\cite{candes2010matrix}, leveraging the low-rank nature inherent to multi-photon videos. Crucially, instead of considering each frame of measurement in isolation, NORA considers the measurements as a whole across both its spatial and temporal components. Throughout the entire pipeline, there is no need for complicated optical components or \textit{a-priori} information about the imaging target nor extensive computational resources.

We demonstrate the capabilities of the NORA pipeline by utilizing an advanced two-photon microscopy simulator~\cite{song2021neural}.  We show that with NORA, it is possible to achieve 2P recordings with up to 1/20th of the original raster lines even with realistic obstacles such as noise and motion. From the combined design of simple optical changes and holistic data recovery, NORA address a significant bottleneck in increasing the frame rate of two-photon microscopes, enabling possible applications with larger FOVs and faster indicators.

\section{Background} \label{background}

\textbf{Multi-photon imaging:}
Multi-photon imaging leverages the nonlinear nature of photon absorption to improve optical resolution deeper into scattering tissue. Rather than illuminating at the wavelength of the fluorescent protein's band gap, light at half the energy (for 2P imaging, lower for three-photon (3P)) is used. Multi-photon excitation provides highly-localized excitation, which enables light collection from more specific locations of the tissue~\cite{luu_more_2024}. Combined with fast galvanometer raster scanning technologies, this excitation "point" can be swept across the tissue to sample across a specified FOV. Two-photon technology is highly reliant on raster scanning, unlike one-photon imaging (1P) where the full FOV is imaged with a CCD array. Raster-scanning is a primary factor behind 2P imaging's high spatial resolution, but the time needed to raster-scan the full FOV also poses a frame rate bottleneck, limiting its use in capturing ultra-fast neural dynamics. A straightforward way to circumvent this bottleneck would be to reduce the number of scanning lines, effectively increasing the frame rate, however restricting the raster-scanning sacrifices spatial resolution for temporal resolution. Addressing the spatial-temporal resolution tradeoff is thus a major motivation for novel multi-photon imaging systems.

\textbf{Computational imaging}:
Computational imaging is an approach for co-designing a combination of novel optics with post-processing algorithms by which the optics encodes the incoming light and the algorithm decodes the measured, encoded samples into a reconstructed image. Computational imaging relies on a modeling the optics encoder through a \textit{forward model}
\begin{gather}
    \bm{y} = g(\bm{x}) + \bm{\epsilon},
\end{gather}
that describes how the $N$-pixel image $\bm{x}\in\mathbb{N}$ is encoded into $M$ samples $\bm{y}\in\mathbb{R}^M$, where here $\bm{\epsilon}\in\mathbb{R}^M$ represents measurement noise. $g(\bm{x})$ can be, for example, the mixing of light through a scattering medium (e.g., the diffuser-cam \cite{antipa_diffusercam_2017}), a stereoscopic projection~\cite{song2017volumetric}, or a tomographic projection~\cite{kazemipour2019kilohertz}.
Recovering the image $\bm{x}$ from the coded measurements $\bm{y}$ is accomplished through an \textit{inverse problem} that finds that image that best explains the measurements through the forward model
\begin{gather}
    \widehat{\bm{x}} = \arg\min_{x} \left\| \bm{y} - g(\bm{x})\right\|_2^2 + \mathcal{R}(\bm{x}),
\end{gather}
where the first term penalizes mismatches to the data and $\mathcal{R}(\bm{x})$ is a regularization term that promotes known statistics of the image $\bm{x}$. For example, $\mathcal{R}(\bm{x})$ can enforce smoothness though the so-called total variation norm ~\cite{rudin_nonlinear_1992}. Regularization becomes especially important when the optics forward model is not well posed, i.e., many possible images $\bm{x}_k$, for $k=1,...,K$ can re-create the same data equally well: $\left\| \bm{y} - g(\bm{x}_{k_1})\right\|_2^2\approx \left\| \bm{y} - g(\bm{x}_{k_2})\right\|_2^2$. The regularization serves as a ``tie-breaker'', making the problem well posed again by further restricting the solution to best align with the expected statistics of the target data.

One major benefit of computational imaging  is its flexibility in designing novel encoding schemes. This is particularly powerful for systems like two-photon microscopy with sampling rate bottlenecks. Some approaches try to mitigate this limitation by elongating the PSF along one axis to integrate multiple points into a single measurement. Examples include the SLAP microscope, which uses line foci instead of point foci to generate tomographic projections~\cite{kazemipour2019kilohertz} and adaptive line-excitation~\cite{li_high-speed_2024} where the PSF is elongated along the fast scan direction. However, these existing methods rely on random-access and require prior structural knowledge of the tissue, restricting generalizability. Furthermore, these priors make these techniques more susceptible to motion artifacts.

\textbf{Compressive sensing and matrix completion}: Compressive (or compressed) sensing (CS)~\cite{romberg2008imaging} and matrix completion~\cite{candes2010matrix} recover signals or data from limited measurements by leveraging underlying structure (sparsity or low rank). In compressive sensing, we recover a $K$-sparse vector $\bm{x} \in \mathbb{R}^N$, i.e, $K \ll N$ elements of $\bm{x}$ are non-zero, from undersampled linear measurements $\bm{y} = \bm{A}\bm{x} + \epsilon$. Here $\mathbf{y}\in \mathbb{R}^M$ for $M<N$. Classical CS theory guarantees the recovery of $\bm{x}$ from $\bm{y}$ when, e.g., $\bm{A}$ satisfies the Restricted Isometry Property (RIP)~\cite{candes2008restricted} or Null-Space Property (NSP)~\cite{mansour2017recovery} and produces enough measurements, typically of the form $M\geq CK\log(N)$ for a constant $C$, i.e., $\bm{y}$ can be much smaller than $\bm{x}$ and still recover it accurately. Often the properties over $\bm{A}$ not hold absolutely, but with high probability if $\bm{A}$ is a random sensing matrix. While CS works well for certain distributions over $\bm{A}$, e.g., \textit{i.i.d.} Gaussian or random Fourier, it does not work well for others, e.g., 1/0 random Bernoulli elements.

Similar to CS, matrix completion recovers a low-rank ($R$) matrix $\bm{X}\in\mathbb{R}^{N\times T}$ from a subset of its entries, leveraging the low-rank property instead of sparsity. $\bm{X}$ can be recovered via nuclear norm minimization, a convex relaxation of rank minimization:
\begin{equation}
\widehat{\bm{X}} = \arg\min_{\bm{X}} \|\mathcal{A}(\bm{X}) - \bm{y}\|_2^2 + \lambda \|\bm{X}\|_{\ast},
\end{equation}
where $\bm{y}$ are the observed entries and $\mathcal{A}(\bm{X})$ selects the corresponding entries of $\bm{X}$. For matrix completion, similar recovery conditions guarantee accurate estimation of $\bm{X}$
\begin{gather}
M\geq CR\left(N+T\right)\log\left(NT\right).
\end{gather}
Here the number of samples $M$ required to recover $\bm{X}$ depends on the latent dimensionality $R(N+T)$, rather than the full number of samples $NT$, leading to a much higher sampling efficiency~\cite{candes2010power, candes2010matrix}.

\section{NORA System Design}

\begin{figure}[t]
    \centering
    \includegraphics[width=0.95\textwidth]{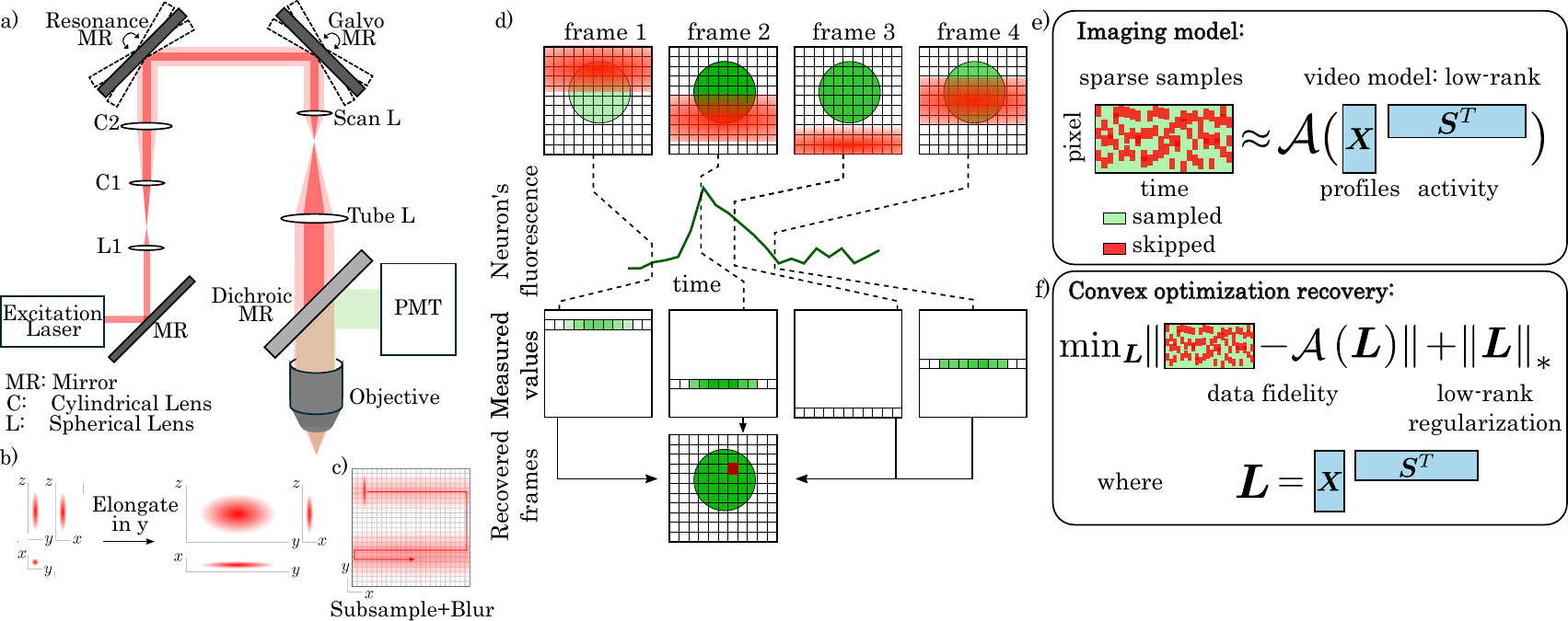}
    \caption{\small Overview of NORA design. a) the proposed design for the NORA microscope, which is the standard two-photon microscope with the addition of two cylindrical lenses to achieve an elongated PSF. b) Elongated PSF shape used in the NORA design. c) Example sampling pattern depicting the blur-and-subsample approach. d) Schematic depicting the recovery process. A single pixel is integrated in multiple wide-line scans. The combined information over multiple frames, along with the correlations between frames, provides enough information to triangulate the value of the single high-resolution pixel at all frames. e) The imaging model quantifies the optical path as a blur operator applied to a low-rank fluorescence video matrix, followed by a sub-sampling operator. Importantly, all operations in the forward model are linear. f) The recovery process solves a nuclear-norm regularized least-squares optimization. This optimization finds the video that both matches the observed samples and the \textit{a priori} model of fluorescence video being low-rank.  }
    \label{fig:content_background}
    \vspace{-0.6cm}
\end{figure}

At a high level, our design is a combined modification to the optical scanning schedule combined with an algorithmic back-end that reconstructs the partially sampled data (Fig.~\ref{fig:content_background}). The optical front-end design seeks to minimize the time spent per frame by reducing the number of lines acquired. This approach leverages the fast resonant galvos used in typical 2P designs while still reducing the total sampling needed to sample each image. The back end is based on matrix completion, an estimation approach that can leverage correlations in both space and time at once.

 \textbf{Data Acquisition}
The Neuroimaging by Oblong Random Acquisition (NORA) data acquisition works as thus: For every frame, an elliptical excitation beam randomly line-scans a fraction of the entire field-of-view. The sampled lines are intentionally selected to be different at each frame, with the goal diversifying the information captured between concurrent frames (by avoiding imaging the same locations). The elliptical PSF is elongated only in the slow-scan direction, covering the area that is equivalent to several raster-lines in a traditional two-photon microscope. Crucially, while the excitation beam is elongated along the slow scan axis, it is not elongated to the point that it accounts for all of the unsampled lines. Furthermore, lines are not spaced out to prevent overlaps, meaning that lines closer than the elongation length may both be collected in the same frame. Each frame will thus have unsampled areas, parts of the field-of-view that are not captured at that time point by any line.

The random line subsampling directly enables the imaging speedup, as the time to acquire each frame is proportional to the number of lines scanned. However, at the subsampling scale of NORa, subsampling with a standard gausssian PSF is insuffient due to how much of the image remains unsampled. When the number of sampled lines become increasingly small, a simple subsampling-only system would pose a significant limit on the total amount of information that could be acquired. The oblong PSF serves to partially account for unsampled lines by illuminating the lines directly surrouunding each scanned lines, collecting fluorescence signal from tissue that otherwise would have been passed over. The signal is collected all at once, effectively summing together the neighboring lines, weighted by the elliptical PSF intensity decay. The ability to acquire signals from multiple lines simultaneously through the elliptical PSF allows for a further reduction in the number of raster lines required per frame while still effectively capturing that frame's information from our target.

Mathematically we can consider the observed measurements as coming from a forward model that consists of two stages, each of which is well modeled by a linear operator. If we denote the ``true'' image at time $t$ that would have been acquired with a slower scan as $\bm{X}_t$, the result of the NORA optical design can be described by first applying a blur operator through a convolution as $\bm{b}\ast\bm{X}_t$, where $\bm{b}$ is the two dimensional representation of the elliptical point-spread function. We can consolidate notation by vectorizing the frame at time $t$ as a sinlge vector $\bm{x}_t$ rather than a 2D image, and we can define an equivalent matrix $\bm{B}$ such that the blurring operation is a simple matrix multiple $\bm{B}\bm{x}$\footnote{We note that $\bm{B}$ has a complex 2-level block circulant structure, however the form of $\bm{B}$ is not important as all that matters is that we can apply the blurring operation to any image. This application can be done in the image domain as a convolution, and the notation here is simply for mathematical efficiency.} Given the blurred version of the image, the measurements are simply the selection of some of the resulting values. We denote this operation as $\bm{S}$, which is an $M\times N$ matrix of zeros and ones where each row has a 1 only in the column corresponding to a collected measurement. The recorded data for image $t$, which we denote $\bm{y}_t$ is thus
\begin{gather}
\bm{y}_t = \bm{S}_t \bm{B} \bm{x}_t + \bm{\epsilon},
\label{eqn:forwardT}
\end{gather}
where $\bm{\epsilon}$ represents the observation noise. Note that $\bm{S}_t$ is time-dependent as the samples collected for each image in the video sequence are different.

The resulting subsampled-and-blurry measurements will still have missing information from the target if each frame is seen in isolation,
despite the elliptical beam ameliorating the amount of information lost.
This loss is mitigated by the design choice of selecting different randomly selected lines at different frames. Thus areas not captured on a frame at time $t$ are likely to be captured at frames in time $t-1$, $t+1$, etc.
It is then up to the reconstruction algorithm to leverage these diverse measurements across multiple frames to not just deblur the lines that were collected with the spread PSF, but to fill in the missing areas  using information shared \textit{between} frames.

\textbf{Computationally Recovering Missing Measurements:}
Given the subsampled measurements, NORA recovers the video by leveraging the strong spatial and temporal correlations in two-photon microscopy videos of neural activity. Specifically, NORA uses the specific mathematical model that fluoresence microscopy data is low-rank. Rank refers to the number of linearly independent rows or columns in a matrix. If a matrix has a rank significantly lower than its dimensions, then it is referred to as low-rank. One of the useful properties of low-rank matrices is that they can be decomposed from a high-dimensional form into much smaller matrices (Fig.~\ref{fig:content_background}e is one such example). The more compact representation of the data makes low-rankness particularly useful in the case of big datasets or large language models, where small and simple approximations are much needed~\cite{bell_lessons_2007, hu_lora_2021}.

In the context of microscopy data, if we consider all frames as vector ($\bm{x}_1$, $\bm{x}_2$, ...) concatenated into a pixels-by-time matrix ($\bm{X} = [\bm{x}_1, \bm{x}_2, ..., \bm{x}_T] \in \mathbb{R}^{N \times T}$), we can safely assume that the rank of $\bm{X}$ is $R\ll \min(N,T)$. To understand the low-rank assumption, consider a sequence of frames: In any sequence, any particular frame is likely to be similar to its preceding frames. Specifically, each frame should capture the same anatomical object, simply with slightly different fluorescence levels per object at each time (refer to the first row of Fig.~\ref{fig:content_background}.d, where the profile is simply changing in intensity). The shared information between frame greatly reduces the effective complexity of the data, enabling a low-rank model for fluorescence microscopy data. This has been successfully done for denoising and segmentation \cite{buchanan_penalized_2018, mukamel_automated_2009, benisty_data_2022}, but utilizing low-rank to recover the $N\times T$ video from fewer observations is still under-explored.

The NORA recovery algorithm utilizes a low-rank prior to reconstruct a full video from under-sampled measurements. To understand the algorithm, we first recall the mathematical representation of our data acquisition pipeline and extend the per-frame forward model (Equation~\eqref{eqn:forwardT}) to a model over a sequence of frames.
\begin{gather}
\bm{Y} = \mathcal{A}(\bm{X}) + \bm{E} = \bm{S}(\bm{B} \bm{X}) + \bm{E}, \label{eqn:forward}
\end{gather}
where here $\bm{S}(\cdot)$ is the linear sampling operator that returns for the $k^{th}$ column (i.e, the blurred $k^{th}$ frame) the samples (lines) taken for that frame, and $\bm{E}$ is the $M\times T$ measurement error matrix. Note that here $\bm{S}(\cdot)$ is not a matrix but an operator since each frame is sampled with a different pattern. Thus, while $\bm{S}(\cdot)$ is still linear, it cannot be written as a single matrix that can be applied to all frames, as $\bm{B}$ can. $\bm{Y} \in \mathbb{R}^{M \times T} $ denotes our observed measurements, where each column is the set of $M$ blurry measurements selected at that frame. $\bm{X} \in \mathbb{R}^{N \times T} $ likewise represents the original imaging target across time.

With a model of how the observed measurements are acquired, it becomes possible to recover the imaging target (a full, non-blurry video) by inverting the forward model. However, this inversion is a non-trivial problem. Despite the model being linear, using a direct pseudo-inverse is not possible due to the fact that $MT<NT$. The under-determined nature of the inverse problem means that an infinite number of solutions exist. To compensate for the incomplete measurements, we use the prior knowledge of the low-rank model.

We frame, as in much of the computational imaging and inverse problem literature, the high-level goal of determining an estimate $\widehat{\bm{X}}$ that satisfies both our observed measurements and our \textit{a priori} model of low-rank fluorescence video.
Imposing this a-priori constraint enables us, as it has had in many other applications, the recovery of $\bm{X}$ despite the ill-conditioned form of the forward model. Mathematically we find the estimate $\widehat{\bm{X}}$ through the nuclear-norm formulation of low-rank matrix completion (Fig.~\ref{fig:content_background}f), i.e., we optimize the cost function
\begin{gather}
    \widehat{\bm{X}} = \arg\min_{X} \left\| \bm{Y} - \bm{S}(\bm{B}\bm{X})\right\|_F^2 + \lambda\|\bm{X}\|_*, \label{eqn:inverse_NORA}
\end{gather}
where the first term (Frobenius norm) is the data fidelity term that enforces the estimate to match the observed samples, and the second term is the nuclear norm (defined as the sum of singular values) that prioritizes low-rank solutions. $\lambda$ is a parameter that trades off between these terms. Solving this optimization has been well studied in the literature. For our implementation, we selected an efficient first-order optimizer that is readily available \cite{becker_templates_2011}.

In addition to efficient optimizers, the nature of nuclear norm optimization for linear forward models also permits theoretical understaning of the expected reconstruction accuracy. Specifically, for NORA imaging we can derive the following guarantees:
\begin{theorem} \label{thm:theorem1}
    Consider the NORA imaging procedure where $L'$ samples are taken at every frame for $T$ frames, i.e., the total number of measurements is $M=TL'$. Assume $TN \geq M$, $N>R$, and $N,T>O(1)$. When
    $$ M \geq C\beta R(T\mu_b^2 + N)\log^2(NT). $$
    Then with probability at least $1-O\left((TN)^{-\beta}\right)$ the solution to Equation~\eqref{eqn:inverse_NORA} produces an estimate $\widehat{\bm{X}}$ of the rank-$R$ matrix $\bm{X}$ with an estimation accuracy bounded by
    $$ \left\|\bm{X}-\widehat{\bm{X}}\right\|_F \leq \left( 4\sqrt{ \min(T,N)\left(2NT + M\right)/M }\right)\epsilon, $$
    where $\epsilon$ is the per-pixel noise error and $\mu_b^2$ measures the coherence of the left singular vectors of $\bm{X}$ and the PSF $\bm{b}_n$ at different locations $n$: $ \mu_b^2 = (N/R)\max_n \|\bm{U}^T\bm{b}_n\|_2^2$.
\end{theorem}

This Theorem demonstrates that for low-rank data, the sampling rate can be quite low, i.e., proportional to the rank times the sum of the matrix dimensions, rather than the full number of elements in the video matrix. This result aligns with past matrix completion results, and is proven using a dual certificate approach (see Supplement for the full proof). One key element of our theory is that the coherence of the PSF shape and the left singular vectors of $\bm{X}$, $1\leq\mu_b^2\leq N$, limits the recovery. In particular, if the principal components ``look like'' the PSF, then $\mu_b^2\rightarrow N$, resulting in a trivial $M\geq NT$ and requiring full sampling. Thus a slight blur, which does not have the appearance of single cells or curved dendrites, is a reasonable choice for the NORA design.

Our easy-to-implement design eliminates many of the difficulties associated with high-speed two-photon microscopy such as cost and expertise. Both the data acquisition and reconstruction aspects of NORA solely rely on simple changes to pre-existing structures. The hardware design of the NORA system is remarkably similar to the traditional two-photon system, only requiring an addition of a cylindrical lens pair to achieve the elongated point spread function~\cite{li_high-speed_2024}. Likewise, the reconstruction does not rely on algorithmically complicated and expensive steps such as model pre-training or gathering large training datasets. These two components taken together removes major barriers towards ultra-fast microscopy.

\begin{figure}[t]
    \centering
    \includegraphics[width=0.95\textwidth]{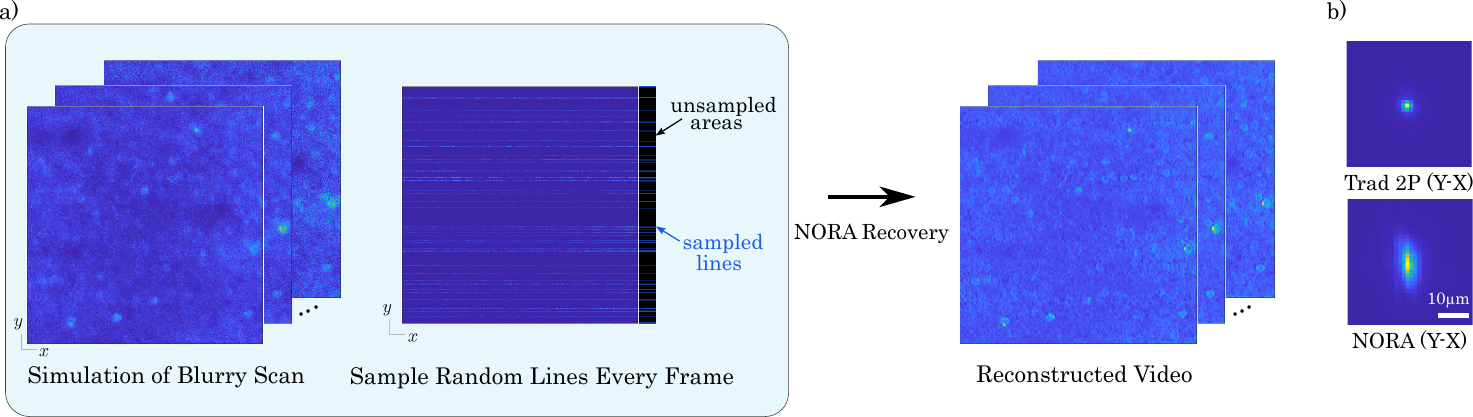}
    \caption{\small NAOMi simulation of NORA imaging. a) NAOMi simulates a full 3D volume of tissue \textit{in silico}, which is then imaged with the desired PSF through an optics model that includes tissue-based aberration and photonic variability. The result is a blurry image of the tissue at each frame. Biophysical models of neural activity ensure realistic changes in fluorescence for the simulated neurons and processes, as well as for neuropil. Frame-specific sampling then selects the per-frame lines to keep, which are then inverted through the matrix completion recovery to obtain an estimate of the videos. Simulations through NAOMi enable the recovered video to be compared to ``standard'' imaging obtained through simulating a video sequence on the same volume and activity, but with a typical PSF and no subsampling. b) Example PSF used in the NAOMi simulation (bottom) compared to a traditional 2P PSF (top).}
    \label{fig:simulation}
    \vspace{-0.6cm}
\end{figure}

\section{Results}

\textbf{Validation through Simulation:}
We validate our design through simulation of our NORA optical path in a biophysical simulation suite followed by reconstruction using the matrix completion approach (Fig.~\ref{fig:simulation}). We simulated the subsampled-and-blurry measurements through the NAOMi simulator (MIT license), a MATLAB-based two-photon microscopy simulator that can create highly-realistic movies of neural tissue with full ground truth ~\cite{song2021neural}. It is important to emphasize that in our simulation, we do not simply take pre-recorded videos and apply the forward model $\bm{S}(\bm{B})$. Rather, NAOMi simulates the full 3D point-spread function and imaging pathway, inclusive of motion, sensor-realistic noise, etc. Using these realistic simulations, we generated a 400 $\mu$m by 400 $\mu$m by 100 $\mu$m volume of tissue mimicking the anatomy of mouse layer 2/3 in area V1 at 250~$\mu$m depth with GCaMP6f. The volume was fully scanned with an elongated PSF (FWHM of  1.15 $\mu m$ by 3.35 $\mu m$ by 6.99 $\mu m$) at 120 mW power and 30 Hz frame rate over 1000 frames. The resulting blurry video was then subsampled by taking random columns form each frame.

We simulated scans with realistic nuisance variation, including noise and motion, as validated in the original NAOMi paper. We tested different levels of subsampling by undersampling the same video at 10X, 15X and 20X (i.e., 1/10th, 1/15th, and 1/20th of the total lines, respectively). Moreover, while we included tissue motion to simulate the most realistic settings, we further tested the impact of motion by comparing the results to the same video generated with the motion ``turned off''.

We developed the reconstruction algorithm in MATLAB\footnote{Code for both simulation and reconstruction will be released with the publication of this paper.}, using the the Templates for First order Conical Solvers (TFOCS) library (BSD 3-clause license) to solve a modified matrix completion problem~\cite{becker_templates_2011}. TFOCS offers several advantages, computational efficiency from bring a first-order method and the ability to use implicit functions. These advantages greatly mitigate the computational challenges of recovering entire video sequences. All computational work was performed on an Intel Core i7 8700k CPU desktop with 64GB RAM. Every 500 frames of reconstruction took $\approx$2 hours.

\begin{figure}[t]
    \centering
    \includegraphics[width=0.95\textwidth]{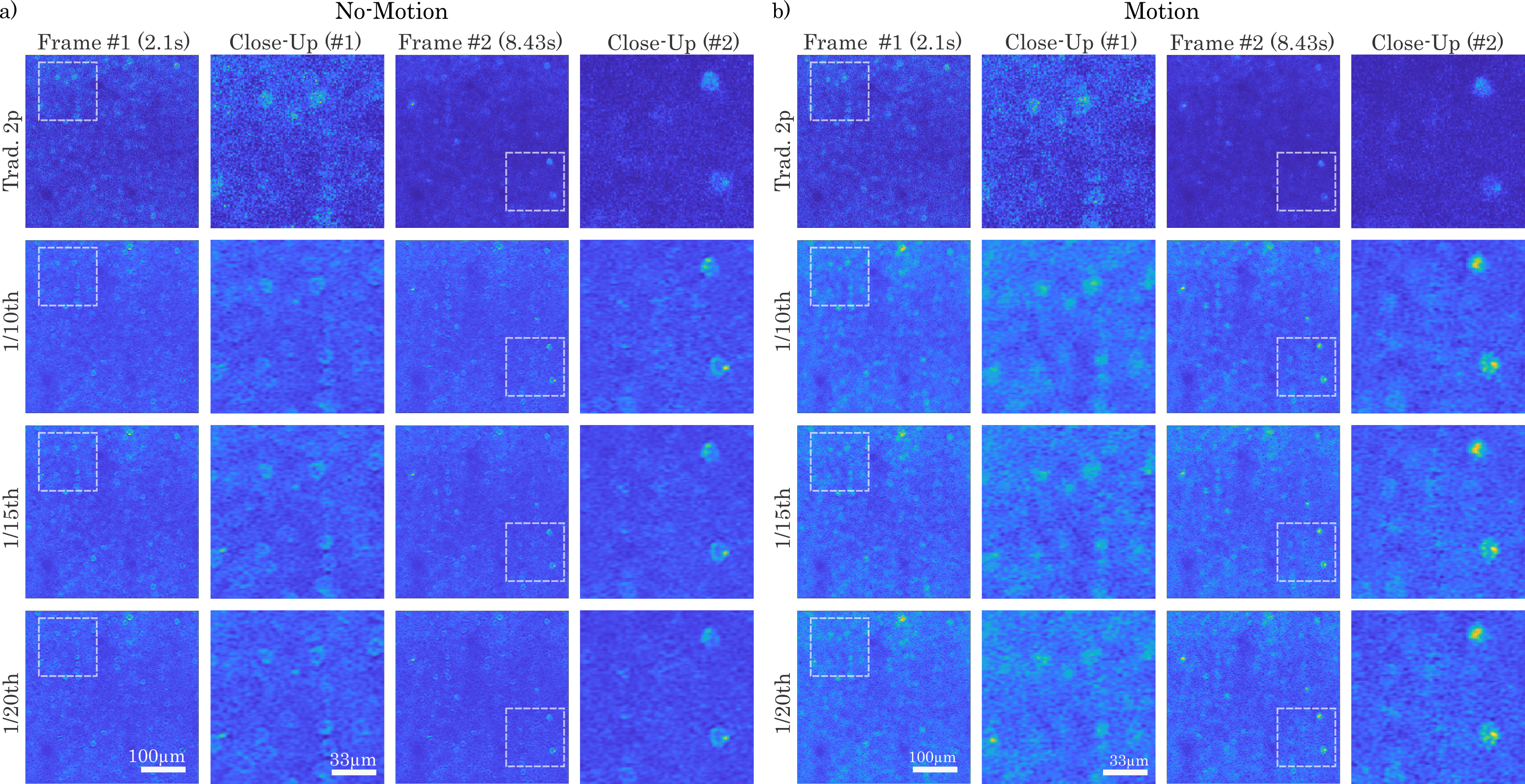}
    \caption{\small  Spatial recovery results for NORA imaging using NAOMi simulations with no motion (a) and with motion (b). (a) and (b) both show example single frames and insets from using (from top to bottom) a traditional 2P PSF, NORA imaging at 10X, 15X, and 20X speedups (1/10$^{th}$, 1/15$^{th}$, and 1/20$^{th}$ of the lines per frame, respectivly). While (a) maintains a slightly higher level of spatial resolution, individual cells are still clearly identifiable up to NORA at 20X for both cases. }
    \label{fig:spatial_recovery}
    \vspace{-0.6cm}
\end{figure}

\begin{figure}[t]
    \centering
    \includegraphics[width=0.95\textwidth]{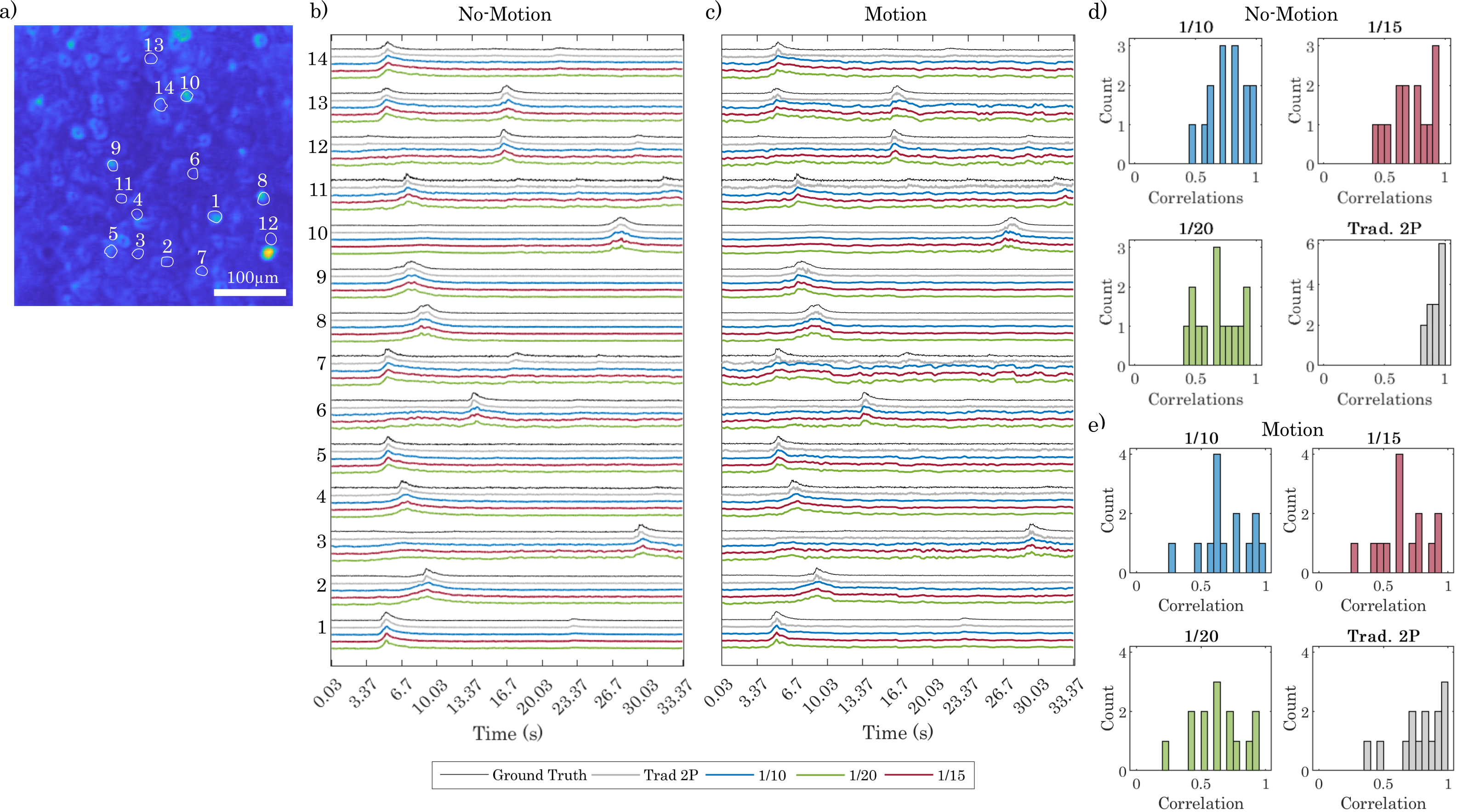}
    \caption{\small  Temporal activity recovery results for NORA imaging. a) Standard deviation image of the median filtered reconstruction with circled ROIs representing individual neurons with significant activity, as identified from NAOMi simulator. b) Time traces from for the ROIs in (a)  for simulations without motion.
    c) Same as (b) but for simulations with motion. In both (b) and (c), activity is preserved up to 20X speedups. d,e) histograms of correlations between each imaging method and the ground truth time-traces for the cases with and without motion. While traditional 2P imaging has the highest correlations, the drop in correlations are minimal, indicating well-preserved neural activity. }
    \label{fig:temporal_recovery}
    \vspace{-0.8cm}
\end{figure}

\textbf{Video recovery from NORA samples:}
Following the NORA pipeline, we simulated the NORA subsampling and blurring for different subsampling ratios: 1/10th, 1/15th, and 1/20th of the FOV. In Figure~\ref{fig:spatial_recovery} we highlight example frames from the NORA-reconstructed videos ($\lambda$ set manually; see Supplement) using frames from a fully-sampled Gaussian PSF scan (traditional 2P) as comparison. We do not include any denoising to highlight the base reconstruction capabilities of NORA.

Despite the extreme subsampling, the NORA reconstructions recover the frames to a visual quality on par with the traditional 2P scan. The reconstruction only shows complete frames of activity without subsampling artifacts and bands of missing information. NORA also remarking maintains a surprising amount of consistency between the reconstructions from different subsampling ratios. As seen through the example frames, there are no major differences between the 1/10$^{th}$ subsampling reconstruction and the 1/20$^{th}$ subsampling reconstruction.

Motion can increase the rank of a video matrix thus violating the low-rank assumption. Specifically, to test the robustness of NORA to rigid motion, we leveraged the ability of NAOMi to simulate both line-by-line and rigid motion. Despite these concerns, the addition of motion only had a minor affect on the visual quality of the reconstructed videos (Fig.~\ref{fig:spatial_recovery}a,b). While some details, e.g., reconstruction sharpness, is affected, the relevant information that needed for ROI extraction, e.g., the location and size of each neuron, are successfully recovered in despite motion and 
high levels of signal recovery are still possible up to 20X speedups. This suggests that although motion does increase the effective rank, the increase is not to the extent that the low-rank model no longer holds.

While the reconstructed videos demonstrated good  image recovery, the core information that is critical in neuroimaging applications  is the temporal activity of individual ROIs in the FOV. If the neural fluorescence fluctuations are distorted, then the imaging method is not well suited to scientific imaging. Conversely, if the images are noisy but the activity is preserved, the imaging method can be used. To examine the time traces, we compared the ground truth traces from NAOMi with the time-traces extracted from ROIs in the NORA reconstructions. We extracted the ROI traces by masking  reconstructed videos with the ground truth spatial profile provided by NAOMi, i.e., the Profile-Assisted Least Squares (PALS)~\cite{song2021neural} (Fig.~\ref{fig:temporal_recovery}a), removing possible confounds of algorithmic-specific errors, such as false positive transients~\cite{gauthier2022detecting}. To reduce noise, specifically extreme values from the Poisson-Gaussian nature of the PMT noise, and focus on signal recovery, we further median filtered all videos with a small 3D window (9x9x9 pixels) before extracting traces. We further compared the NORA recovered ROI traces to traces from the traditional 2P simulation scan. This comparison provided a reference for how well a typical system would have captured the ground truth activity. Traces are shown for 33.33s of simulated recording (Fig.~\ref{fig:temporal_recovery}), both for simulations without motion (Fig.~\ref{fig:temporal_recovery}b,d) and with motion (Fig.~\ref{fig:temporal_recovery}c). 

The traces extracted from the reconstructed volumes are largely in line with the traces from the ground truth for both with and without motion (Fig.~\ref{fig:temporal_recovery}b,c). Traces such as \#1, \#3, and \#6 show the consistency in the spike between the ground truth and the reconstruction. In cases like trace \#7 (Fig.~\ref{fig:temporal_recovery}c), we see that the traditional 2P traces also fail to match the ground truth, suggesting that the failure is due to a general scanning issue rather than NORA.

One important question we consider is the general trend of the reconstructions: do most traces consistently align with the ground truth? To verify the overall alignment of the extracted traces with the ground truth, the distribution of the correlation between the ground truth traces and the extracted traces are plotted in Figure~\ref{fig:temporal_recovery}d,e. Even in the simulation condition that includes motion, all traces from the reconstructed volumes show a skew toward the left, indicating a general trend of high correlation with the ground truth despite the inclusion of noise and motion in the measurement process.

\section{Discussion}

In this work we present NORA: a framework for fast imaging of raster-scanning two-photon imaging. NORA is based on a subsample-and-blur optical design combined with a matrix-completion recovery approach. The advantages of our system are 1) the simplicity of the optical design, 2) the use of both spatial and temporal statistics for video recovery, and 3) theoretical guarantees on recovery. Moreover, NORA uses generic low-rank statistics and thus does not require any training data or model fitting. 

One major feature is NORA's robustness to both extreme subsampling and motion. Given that motion increases the effective matrix rank by mis-aligning pixels across frames, we anticipated requiring additional algorithmic improvements to address motion. Despite this rank increase, however, and we were able to accurately reconstruct the video from sub-sampling levels up to a factor of 20X. This successful recovery highlights that NORA is capability of handling realistic data challenges.  

In our approach we focused on traditional  matrix reconstruction via nuclear norm minimization, which has been highly successful in other domains~\cite{candes2010matrix,charles2017distributed,ahmed2014blind}. This approach moves from thinking of compressive microscopy recovery as per-image frame-by-frame recovery to using the full set of spatial and temporal statistics. Moreover, low-rank priors are powerful enough that reconstruction is possible without strong trained priors learned by recent deep learning approaches. We thus bypass the need for training data, and remove the possibility of realistic hallucinations that can be difficult to diagnose. Furthermore, NORA has the capacity to generalize to new tissue samples, unlike deep learning approaches that face significant out-of-distribution challenges~\cite{mishne2024deep}.

Interestingly, many ROI detection algorithms are based on low-rank matrix factorization~\cite{charles2022graft,mishne2019learning,giovannucci2019caiman} (see~\cite{benisty_data_2022} for a recent review). Possible future direction could thus consider merging ROI detection with  NORA reconstruction, saving computational effort and reducing possible artifacts from multiple stages of analysis. Here, however, we focus on reconstruction as in enables users to apply any ROI detection algorithm. Moreover, different cell types (e.g., astrocytes) or imaging targets (e.g., dendrites) require specialized ROI detection~\cite{mi2024fast,charles2022graft}, 
further justifying reconstruction as a first step. 

NORA's simple, practical design can enable many labs to build an ultra-fast microscope with with minimal additional parts and effort. 
We envision two possible use cases for NORA: fast imaging of a single FOV, and interleaved imaging of multiple FOVs. In the former, a single FOV is sampled as fast as possible, using the full framerate to acquire signals faster than calcium dynamics (e.g., voltage~\cite{evans2023positively} or glutamate~\cite{aggarwal2023glutamate} indicators). This use case has a many applications in studying neural interactions based on fast spike timing. In the latter, random access refocusing can flexibly select multiple (10-20) independent FOVs over the optically accessible volume without specific, required geometrical configurations (e.g., tiling in depth). The second use case opens up the ability to study the network-level activity of multiple brain areas at the single neuron resolution. 

\textbf{Limitations:} The physical design proposed for NORA only requires simple optical components. One aspect, however, that may differ from simulation is the random sampling implementation. In simulation, each frame's measurements consists of randomly selected columns. A physical system would require the galvanometer to modulate its acceleration to accommodate the varying distances between samples, which could cause problems such as jerk and slow down the system. A preferable sampling would consist of uniform distances between each sample, such as evenly-spaced columns of measurements, to allow for consistent speed between measurement column. 
The flexibility of NORA allows for this type of sampling, as long as the measurements are sufficiently different across frames. 
Our work largely focused on the validation of NORA through theoretical guarantees and biophysical simulation. Validation in practice requires extensive simultaneous imaging that, while expensive and time-intensive, represent exciting next steps. Another limitation is the current reconstruction run-time, which we are actively working to improve with better hardware (GPUs) and optimized software. 

\section*{Acknowledgments} EW and JY were funded by NIH award R01EB034272. ASC was in part funded by CZI award CP2-1-0000000704. 

\bibliographystyle{plain}
\bibliography{refs}

\clearpage

\appendix
\section{Appendix: Implementation Details}

Here we provide additional information on parameters necessary to implement the NORA reconstruction. The implementation builds on the templates provided in the TFOCs library~\cite{becker_templates_2011}, and likewise utilize their formulation of the nuclear norm basis pursuit denoising (BPDN) problem. Overall, the NORA reconstruction is based on the regularized version of the nuclear (BPDN) problem:
\begin{equation}
\widehat{\bm{X}} = \arg\min_{\bm{X}} \|\mathcal{A}(\bm{X}) - \bm{y}\|_2^2 + \lambda \|\bm{X}\|_{\ast},
\end{equation}
where $\bm{X}\in\mathbb{R}^{N\times T}$ is a low-rank ($R$) matrix that we wish to estimate, $\bm{y}\in\mathbb{R}^{M}$ are the observed entries and $\mathcal{A}(\bm{X})$ is the feed-forward model that describes mathematically how the data in $\bm{X}$ is projected into our measurements $\bm{y}$. The choice in value for $\lambda$, the regularizer parameter, determines the relative strength of the low-rank prior.

The TFOCs implementation essentially seeks to solve the same problem but translating the problem into a constrained optimization form. This form further includes a smoothing term in order to make the problem better posed, resulting in the following formulation for Nuclear BPDN:
\begin{align}
\widehat{\bm{X}} = \arg\min_{\bm{X}} \quad & \|\bm{X}\|_* + \frac{1}{2}\mu \|\bm{X} - \bm{X}_0\|_F^2 \\
\text{subject to} \quad & \| \mathcal{A}(\bm{X}) - \bm{y} \| \leq \epsilon,
\end{align}
where the second term in the cost function is the smoothing term, $\bm{X}_0\in\mathbb{R}^{N\times T}$ is the optional initial point, $\mu$ is the smoothing parameter, and $\epsilon$ is the noise parameter. TFOCS can run in `continuation' model where the solution is fed back into the optimization as the initial point $\bm{X}_0\leftarrow\widehat{\bm{X}}$ until the solution does not deviate much from the initial condition. This approach requires multiple runs of the optimization, which due to the size of the optimization and related computational run-time we opted to not perform. Instead we used a small, constant $\mu$. For our implementation of the NORA reconstruction, careful selection of $\epsilon$ is essential to guaranteeing successful reconstruction. We used the following values of $\epsilon$ for the reconstructions shown in Figures~\ref{fig:spatial_recovery} and~\ref{fig:temporal_recovery}, which were reconstructed in batches of 500 frames:

\begin{table}[h!]
\centering
\begin{tabular}{lccc}
\toprule
 & \textbf{1/10th} & \textbf{1/15th} & \textbf{1/20th} \\
\midrule
\textbf{Noise, No Motion} & 475  & 375 &325  \\
\textbf{Noise, Motion}    & 475 & 375  & 340  \\
\bottomrule
\end{tabular}
\caption{$\epsilon$ values for simulated reconstructions }
\end{table}

The above values for $\epsilon$ were chosen manually based on reconstruction quality. $\mu$ was kept consistent as $\mu=0.1$ for all reconstructions.

We note that there appears to be a relationship between the number of measurements and the value of $\epsilon$, with $\epsilon$ scaling with increased measurements. Intuitively, this could be due to the smaller number of measurements introducing less variations, allowing for a smaller error tolerance. When implementing NORA, we recommend starting with the given $\epsilon$ values, then scaling depending on the number of frames, size of the FOV, and subsampling ratio.

\section{Appendix: Proof of NORA reconstruction}

This appendix aims to prove Theorem~\ref{thm:theorem1}, which bounds the recovery error of estimating a low-rank video sequence $\bm{X}\in\mathbb{R}^{N\times T}$  from the NORA measurements $\bm{Y}\in\mathbb{R}^{M}$. The basis of the proof rests on earlier work in the matrix recovery literature that have derived recovery guarantees for similar systems using a dual certificate approach certificate approach~\cite{candes2010power,candes2010matrix}. In the dual certificate approach, we use the fact that if there exists a ``certificate'' that satisfies a series of inequalities derived from the KKT conditions, then the low-rank matrix is recoverable from a nuclear norm minimization.

Specifically, the certificate $\bm{Z}$ must satisfy two bounds, one on its projections within the span of the singular vectors of $\bm{X}$, and one on the projection outside of that same span.
To prove that such a certificate exists, it is sufficient to find one such instance of a matrix that satisfies all the proper conditions, i.e., ``proof by construction''.

We start by defining the singular value decomposition of the video sequence $\bm{X}$ as $\bm{S} = \bm{Q}\bm{\Sigma}\bm{V}^T$. To be able to simply define the dual certificate properties that must hold, we next define the projections into the space spanned by the left and right singular vectors $\calP$, and projection into the complement space $\mathcal{P}_{T^{\perp}}$ as
\begin{eqnarray}
    \opcalP{\bm{W}} & = & \bm{Q}\bm{Q}^T\bm{W} + \bm{W}\bm{V}\bm{V}^* - \bm{Q}\bm{Q}^T\bm{W}\bm{V}\bm{V}^T \label{eqn:projdef} \\
    \mathcal{P}_{T^{\perp}}(\bm{W}) & = & (\bm{I} - \bm{Q}\bm{Q}^*)\bm{W}(\bm{I} - \bm{V}\bm{V}^T). \nonumber
\end{eqnarray}

Given these projection operations, we can state the dual certificate as
\begin{align}
    \norm{\opcalP{\bm{Z}} - \bm{Q}\bm{V}^T}_F &\leq \frac{1}{2\sqrt{2}\gamma} \label{eqn:dual1} \\
    \norm{\mathcal{P}_{T^{\perp}}(\bm{Z})} &\leq \frac{1}{2}. \label{eqn:dual2}
\end{align}

Our proof approach is based on the golfing scheme, which was developed and used extensively in prior work~\cite{charles2017distributed,gross2011recovering,candes2011probabilistic,ahmed2014blind}. The golphing scheme aims to construct a certificate $\bm{Z}$ that satisfies the above properties. The golfing scheme is an iterative method that begins with a trivial all-zeros certificate $\bm{Z}$ and continually refines the certificate through an update iteration that produces a sequence of  certificates $\bm{Z}_k$ for $k\in[1,\cdots,\kappa]$ that each are closer to satisfying the desired conditions. The sequence is then shown to converge to a final certificate $\bm{Z}_{\kappa}$ that satisfies both conditions. The golfing iterations are defined as
\begin{align}
    \bm{Z}_k = \bm{Z}_{k-1} + \kappa\calA^T_k\calA_k(\bm{Q}\bm{V}^T - \calP(\bm{Z}_{k-1})), \nonumber
\end{align}
where the operator $\calA(\bm{W})$ represents a single sample of the observation matrix, i.e.,
\begin{equation}\label{eqn:linopdef}
    \calA(\bm{W}) = \mbox{vec}(\dotprod{\bm{A}_n}{\bm{W}}).
\end{equation}

\subsubsection*{Convergence of the golfing scheme}

Using the golfing scheme we need to show that subsequent applications of the iterations converge to a certificate that satisfies the required properties. To show the first property in Equation~\eqref{eqn:dual1} we consider a refined, simpler update iteration used in prior work~\cite{ahmed2014blind,charles2017distributed} deined in terms of a modified certificate $\widetilde{\bm{Z}}_k = \calP(\bm{Z}_k)-\bm{Q}\bm{V}^T$:
\begin{gather}
    \widetilde{\bm{Z}}_k = (\calP - \kappa\calP\calA^T_k\calA_k\calP)\widetilde{\bm{Z}}_{k-1}, \nonumber
\end{gather}

It follows that one must show that this iterative procedure converges, with high probability, to a certificate satisfying the desired dual certificate conditions. The rest of the details in the appendix are dedicated to this demonstration.

\subsubsection*{High level convergence of the dual certificate}

The first condition we need to demonstrate is that the Forbenious norm of the $k^{th}$ iterate, $\widetilde{\bm{Z}}_k$ is bounded with high probability.

by using Lemma~\ref{lem:lemma1} and observing that the Forbenious norm of the $k^{th}$ iterate is well bounded with probability $1-O((NT)^{-\beta})$ by
\begin{eqnarray*}
    \left\| \widetilde{\bm{Z}}_k \right\|_F & \leq & \max_k \left\| \mathcal{P}_T - \kappa\mathcal{P}_T\mathcal{A}^{T}_k\mathcal{A}_k\mathcal{P}_T \right\|\left\| \widetilde{\bm{Z}}_{k-1} \right\|_F \nonumber \\
	& \leq & 2^{-k}\left\| \widetilde{\bm{Z}}_{0} \right\|_F  \nonumber \\
	& \leq & 2^{-k}\left\| \bm{Q}\bm{V}^T \right\|_F \nonumber \\
	& \leq & 2^{-k}\sqrt{R},        \nonumber
\end{eqnarray*}
Where here we use Lemma~\ref{lem:lemma1} so long that $M \geq c\beta\kappa R(\mu^2_b T + N)\log^2(TN)$. As in~\cite{ahmed2014blind,charles2017distributed} we set $\kappa \geq 0.5\log_2(8\gamma^2R)$, which reduces this  bound for the Frobenious norm of $\widetilde{\bm{Z}}_{\kappa}$ is bounded by $\left\| \widetilde{\bm{Z}}_{\kappa} \right\|_F \leq (2\sqrt{2}\gamma)^{-1}$.

Next, we show that the second dual certificate holds:
\begin{eqnarray*}
	{\left\| {\mathcal{P}_{T^{\perp}}\left( \bm{Z}_{{\kappa}} \right)} \right\|} & \leq & \sum_{k=1}^{\kappa} \left\| \mathcal{P}_{T^{\perp}}\left( \kappa\mathcal{A}^{T}_k\mathcal{A}_k\widetilde{\bm{Z}}_{k-1}  \right) \right\|\nonumber \\
	& = & \sum_{k=1}^{\kappa} \left\| \mathcal{P}_{T^{\perp}}\left( \kappa\mathcal{A}^{T}_k\mathcal{A}_k\widetilde{\bm{Z}}_{k-1} - \widetilde{\bm{Z}}_{k-1}  \right) \right\| \nonumber \\
	& \leq & \sum_{k=1}^{\kappa} \left\| \kappa\mathcal{A}^{T}_k\mathcal{A}_k{\widetilde{\bm{Z}}}_{k-1} - \widetilde{\bm{Z}}_{k-1} \right\|         \nonumber \\
	& \leq & \sum_{k=1}^{\kappa} \left\| \kappa\mathcal{A}^{T}_k\mathcal{A}_k\widetilde{\bm{Y}}_{k-1} - \widetilde{\bm{Z}}_{k-1} \right|_F       \nonumber \\
	& \leq & \sum_{k=1}^{\kappa} \max_{k\in[1,\ldots{\kappa}]} \left\| \kappa\mathcal{A}^{T}_k\mathcal{A}_k\widetilde{\bm{Z}}_{k-1} - \widetilde{\bm{Z}}_{k-1} \right\|_F  \nonumber \\
	& \leq & \sum_{k=1}^{\kappa} \frac{1}{2}2^{-k} \nonumber \\
	& \leq & \frac{1}{2}                                  \nonumber
\end{eqnarray*}
We use Lemma~\ref{lem:lemma2} to bound the maximum spectral norm of $\kappa{\mathcal{A}^{T}}_k\mathcal{A}_k\widetilde{\bm{Z}}_{k-1} - \widetilde{\bm{Z}}_{k-1}$ with probability $1-O((TN)^{1-\beta}$. Taking $\kappa \geq \log(LN)$ shows that the final certificate $\bm{Z}_{\kappa}$ satisfies all the desired properties, completing the proof.

\subsubsection*{Useful definitions and theorems}

\begin{theorem}[Matrix Bernstein's Inequality]
    \label{thm:matbern}
    Let $\bm{X}_i\in\mathbb{R}^{L,N}$, $i\in[1,\dots,M]$ be $M$ random matrices such that $\EE{\bm{X}_i} = 0$ and $\norm{\bm{X}_i}_{\psi_\alpha} < U_{\alpha} < \infty$ for some $\alpha \geq 1$. Then with probability $1-e^{-t}$, the spectral norm of the sum is bounded by
    \begin{align}
        \norm{\sum_{i=1}^M \bm{X}_i} \leq C\max\left\{\sigma_X\sqrt{t+\log(L + N)}, U_{\alpha}\log^{1/\alpha}\left(\frac{M U_{\alpha}^2}{\sigma_X^2}\right)(t + \log(L + N)) \right\}, \nonumber
    \end{align}
    for some constant $C$ and the variance parameter defined by
    \begin{align}
        \sigma_X = \maxset{\norm{\sum_{i=1}^M \EE{\bm{X}_i\bm{X}_i^{\ast}}}^{1/2}, \norm{\sum_{i=1}^M \EE{\bm{X}_i^{\ast}\bm{X}_i}}^{1/2} }. \nonumber
    \end{align}
\end{theorem}

where Orlicz-$\alpha$ norm $\norm{X}_{\psi_\alpha}$ is defined as
\begin{align}
    \norm{X}_{\psi_\alpha} = \inf\left\{ y > 0 ~\vert~ \EE{e^{\norm{X}^{\alpha}/y^{\alpha}}} \leq 2 \right\}. \label{eqn:onormdef}
\end{align}

An additional useful lemma from~\cite{tropp2012user,ahmed2014compressive} relates the Orlicz-1 and -2 norms for a random variable and it's square. For completeness, we include this lemma here for easy reference:

\begin{lemma}[Lemma 5.14,~\cite{tropp2012user}]
    \label{lem:onorm1}
    A random variable $X$ is sub-gaussian iff $X^2$ is subexponential. Furthermore,
    \begin{align}
        \norm{X}_{\psi_2}^2 \leq \norm{X^2}_{\psi_1} \leq 2\norm{X}_{\psi_2}^2. \nonumber
    \end{align}
\end{lemma}

\begin{lemma}[Lemma 7,~\cite{ahmed2014blind}]
    \label{lem:onorm2}
    Let $X_1$ and $X_2$ be two sub-gaussian random variables. Then the product $X_1X_2$ is a sub-exponential random variable with
    \begin{align}
        \norm{X_1X_2}_{\psi_1} \leq c\norm{X_1}_{\psi_2}\norm{X_2}_{\psi_2}.   \nonumber
    \end{align}
\end{lemma}

\subsection{Bounding the Frobenius norm}
\begin{lemma} \label{lem:frobound}
The Frobenius norm of the projection $\opcalP{\bm{A}_n}$ is bounded by
\begin{align}
\norm{\opcalP{\bm{A}_n}}_F^2 \leq \norm{\bm{Q}^T\bm{b}_n}_2^2 + \eta\norm{\bm{V}^T\bm{s}_n}_2^2
\end{align}
where $\bm{A}_n = \bm{b}_n\bm{s}_n^T$ represents rank-one samples and blurring, and $\eta = \|\bm{b}_n\|_2^2$ is the total fluorescence integration over the PSF.
\end{lemma}

\begin{proof}
Next, we show that $\norm{\opcalP{\bm{A}_n}}_F^2$,the size of the projection in Frobenius norm is controlled by the number of the samples and blur. Indeed, it follows by algebra that

\begin{align}
\norm{\opcalP{\bm{A}_n}}_F^2 \nonumber
&= \dotprod{\opcalP{\bm{A}_n}}{\opcalP{\bm{A}_n}} \\ \nonumber
&= \dotprod{\opcalP{\bm{A}_n}}{\bm{A}_n} \\ \nonumber
&= \dotprod{\bm{Q}\bm{Q}^T\bm{b}_n\bm{s}_n^T}{\bm{b}_n\bm{s}_n^T} + \dotprod{\bm{b}_n\bm{s}_n^T\bm{V}\bm{V}^T}{\bm{b}_n\bm{s}_n^T} - \dotprod{\bm{Q}\bm{Q}^T\bm{b}_n\bm{s}_n^T\bm{V}\bm{V}^T}{\bm{b}_n\bm{s}_n^T} \\ \nonumber
&= \norm{\bm{s}_n}_2^2\norm{\bm{Q}^T\bm{b}_n}_2^2 + \norm{\bm{b}_n}_2^2\norm{\bm{V}^T\bm{s}_n}_2^2 - \norm{\bm{Q}^T\bm{b}_n}_2^2\norm{\bm{V}^T\bm{f}_n}_2^2 \\
&\leq \norm{\bm{Q}^T\bm{b}_n}_2^2 + \eta\norm{\bm{V}^T\bm{s}_n}_2^2,
\end{align}
where we use here the fact that $\bm{A}_n$ for one sample is rank. 
Therefore,
\begin{align*}
\norm{\opcalP{\bm{A}_n}}_F^2 \leq \norm{\bm{Q}^T\bm{b}_n}_2^2 + \eta\norm{\bm{V}^T\bm{s}_n}_2^2
\end{align*}

\end{proof}

\subsection{Bounding the Operator Norm}
\begin{lemma} \label{lem:lemma1}
Suppose $\calA_k$ is a measurement operator with $\EE{\calA_k^{*}\calA_k} = \frac{1}{\kappa}\mathcal{I}$, where $\bm{A}_n = \bm{b}_n\bm{s}_n^T$ represents rank-one samples. Then with probability at least $1-\frac{1}{(N+L)}$, we have
\begin{equation*}
    \max_{k\in[1,...,\kappa]} \norm{\kappa\calP\calA_k^T\calA_k\calP-\calP}_{op} \leq \frac{1}{2}
\end{equation*}
provided that $M \geq C\frac{R}{M}\left(\mu_b^2T + N\right)\log^2(NT)$ for a large enough constant $C$.
\end{lemma}

\begin{proof}
The goal of this lemma is to prove that with overwhelming probability the operator norm of the function defining the golfing scheme is small, and therefore the certificate norm shrinks by enough to satisfy the dual certificate conditions in $\kappa$ steps. We can achieve the desired bound through a matrix Bernstein inequality. 
We start by noting that since  $\EE{\calA_k^{*}\calA_k} = \frac{1}{\kappa}\mathcal{I}$, the operator whose norm we wish to bound is equivalent to

\begin{align*}
\kappa\calP\calA_k^{T}\calA_k\calP - \calP &= \kappa\calP\calA_k^{\ast}\calA_k\calP - \EE{\kappa\calP\calA_k^{T}\calA_k\calP} \\
&= \kappa\sum_{n\in\Gamma_k}\left(\opcalP{\bm{A}_n}\otimes\opcalP{\bm{A}_n} - \EE{\opcalP{\bm{A}_n}\otimes\opcalP{\bm{A}_n}} \right)
\end{align*}

Next we define the operator $\calL_n(\bm{C}) = \dotprod{\opcalP{A}}{\bm{C}}\opcalP{A}$. Note that the Frobenious norm of this operator is $\norm{\calL(\bm{C})}_F^2 = \dotprod{\calL(\bm{C})}{\calL(\bm{C})} = \dotprod{\calL(\bm{C})}{\opcalP{C}}$ for any $\bm{C}\in M\left(n, \mathbb{R}\right)$, meaning that we can simplify the above as 

\begin{equation*}
\kappa\calP\calA_k^{T}\calA_k\calP - \EE{\kappa\calP\calA_k^{T}\calA_k\calP}
= \kappa\sum_{n\in\Gamma_k}(\calL_n - \EE{\calL_n})
\end{equation*}

\textbf{Computing the variance:} To leverage the Bernstein's inequality and bound the operator norm, we need to compute the variance of the random operator. We observe that the projection $\calP$ is bisectorial and so it suffices to use the symmetry  $\mathcal{L}_n$ to calculate

\begin{equation*}
\kappa^2\norm{\sum_{n\in\Gamma_k}\EE{\calL_n^2} - \EE{\calL_n}^2}
\leq \kappa^2 \norm{\sum_{n\in\Gamma_k} \EE{\calL_n^2}} = \kappa^2\norm{\EE{\sum_{n\in\Gamma_k} \norm{\opcalP{\bm{A}_n}}_F^2\calL_n}}
\end{equation*}

Recall that 
$\bm{A}_n = \bm{b}_n \bm{s}_n^T \in M\left(\mathbb{R},1\right)$ is a rank-1 matrix. Since the point spread function $\bm{b}_n$ is a standard Gaussian shape and $\bm{s}_n$ is a sample vector that has at most $1$ pixel selected, their respective vector norms are $\|\bm{b}_n\|_2^2=\eta$ and $\|\bm{s}_n\|_2^2 = 1$, where $\eta$ is the total integrated fluorescence power the PSF. 
We can then use the bound on the Frobenious norm of $\opcalP{\bm{A}_n}$ from   Lemma~\ref{lem:frobound} and substitute to obtain:

\begin{align*}
\kappa^2\norm{\EE{\sum_{n\in\Gamma_k} \norm{\opcalP{\bm{A}_n}}_F^2\calL_n}}
&\leq \kappa^2\norm{\EE{\sum_{n\in\Gamma_k} \left(\norm{\bm{Q}^T\bm{b}_n}_2^2 + \eta\norm{\bm{V}^T\bm{s}_n}_2^2\right)\calL_n}} \\
&= \kappa^2\norm{\EE{\sum_{n\in\Gamma_k} \norm{\bm{Q}^T\bm{b}_n}_2^2\calL_n + \eta\sum_{n\in\Gamma_k}\norm{\bm{V}^T\bm{s}_n}_2^2\calL_n}}
\end{align*}

To simplify this bound further, we define the following coherence terms:

\begin{eqnarray}
    \mu_b^2 & = & \frac{N}{R}\max \left|\left|\bm{Q}^{\ast}\bm{b} \right|\right|_2^2. \label{eqn:coherencez}
\end{eqnarray}
This term quantifies the average intersection of the fluorescing components in the image with the spatial sampling pattern. Specifically, $\mu_b^2$, measures the average intersection between the PSF at any given location with the objects in the image. At a minimum, $\mu_b^2$ being low requires that the PSF does not have the shape of a single cell. Interestingly, due to the correlations between cells and neuropil, the actual overlap is much smaller as the singular vectors of a typical FOV contain widespread information.

Using the definitions of $\mu_b^2$, we can now rewrite the variance term as
\begin{align*}
\norm{\EE{\sum_{n\in\Gamma_k} \norm{\opcalP{\bm{A}_n}}_F^2\calL_n}}
&\leq \norm{\sum_{n\in\Gamma_k}\EE{\left(\norm{\bm{Q}^T\bm{b}_n}_2^2 + \eta\norm{\bm{V}^T\bm{s}_n}_2^2\right)\calL_n}} \\
&\leq \norm{\sum_{n\in\Gamma_k}\EE{\norm{\bm{V}^T\bm{s}_n}_2^2\calL_n}} + \max\norm{\bm{Q}^T\bm{b}_n}_{2}^2\norm{\sum_{n\in\Gamma_k}\EE{\calL_n}} \\
&\leq \frac{R}{T}\norm{\EE{\sum_{n\in\Gamma_k}\calL_n}} + \frac{R}{N}\mu_b^2\norm{\sum_{n\in\Gamma_k}\EE{\calL_n}} \\
&\leq \left(\frac{R}{T} + \frac{R}{N}\mu_b^2\right)\norm{\sum_{n\in\Gamma_k}\EE{\calL_n}}
\end{align*}

We thus next focus on $\norm{\sum_{n\in\Gamma_k}\EE{\calL_n}}$

\begin{eqnarray}
    \norm{\sum_{n\in\Gamma_k}\EE{\calL_n}} &= & \norm{\sum_{n\in\Gamma_k}\EE{\calL_n}}  \\
    & = & \norm{\sum_{n\in\Gamma_k}\EE{\opcalP{\bm{A}_n}\otimes\opcalP{\bm{A}_n} }} \\
    & \leq & \norm{\mathcal{P}_T }\norm{\sum_{n\in\Gamma_k}\EE{\bm{A}_n\otimes\bm{A}_n }}\norm{\mathcal{P}_T } \\
    & = & \norm{\sum_{n\in\Gamma_k}\EE{
    \bm{b}_n\bm{s}_n\otimes\bm{b}_n\bm{s}_n }} \\
    & \leq & \norm{\sum_{n\in\Gamma_k}\EE{
    \{\bm{b}_n[i]\bm{b}_n[j]\bm{s}_n\bm{s}_n^T \}_{ij}}}_F^2  \\
    & \leq & \frac{1}{T^2}\norm{\sum_{n\in\Gamma_k}
    \{\EE{\bm{b}_n[i]\bm{b}_n[j]}\bm{I}\}_{ij}}_F^2 \\
    & \leq &\frac{T}{T^2L'^2}\norm{\sum_{n\in\Gamma_k}
    \bm{b}_n\bm{b}_n^T}_F^2 \\
    & \leq &\frac{T}{M^2}\norm{\sum_{n\in\Gamma_k}
    \bm{b}_n\bm{b}_n^T}_F^2 \\
    & \leq &\frac{T}{M^2}\sum_{n\in\Gamma_k} \norm{\bm{b}_n\bm{b}_n^T}_F^2 \\
    & \leq &\frac{T}{M^2}\sum_{n\in\Gamma_k} N  = \frac{TN}{M^2}\frac{M}{\kappa} = \frac{TN}{M\kappa}
\end{eqnarray}

Thus,
\begin{eqnarray}
    \sigma_X^2 &\leq &\left(\frac{R}{T} + \frac{R}{N}\mu_b^2\right)\left(\frac{NT}{M\kappa}\right) \\
    &= & \frac{R\left(N+T \mu_b^2\right)}{M\kappa}
\end{eqnarray}


\textbf{Computing the Orlicz norm:} The second term of the Bernsteins inequality requires the Orlicz-1 norm. Since $\calL_n$ is positive semidefinite (PSD), it's a Young function and thus we can compute its Orlicz-1 norm:

\begin{align*}
\kappa\norm{\norm{\calL_n}_F - \EE{\norm{\calL_n}_F}}_{\psi_1}
\end{align*}

We would like to show that with high probability that the projection operator of measurements is small with high probability. In order to do this , we seek to first compute the size of the operator by measuring it in the Frobenius sense and then we bound it using the Matrix Bernstein inequality.

\begin{align*}
\norm{\norm{\calL_n}_F - \EE{\norm{\calL_n}_F}}_{\psi_1} \leq \maxset{\norm{\norm{\calL_n}_F}_{\psi_1} - \norm{\EE{\norm{\calL_n}_F}}_{\psi_1}}
\end{align*}

First, note that the size of $\norm{\EE{\norm{\calL_n}_F}}_{\psi_1}$ can be determined via

\begin{align}
\norm{\EE{\norm{\calL_n}_F}}_{\psi_1} \nonumber
&= \norm{\EE{\Tr{\calP\calA_n\calA_n^T\calP^T}}}_{\psi_1} \\
&= \norm{\EE{\Tr{\calA_n\mathcal{I}\mathcal{A}_{n}^{T} }}} _{ \psi_1} \\ \nonumber
&= \frac{\eta}{T\log\left(2\right)}
\end{align}

Next, to compute $\norm{\norm{\calL_n}_F}_{\psi_1}$ we use the definition of the Orlicz-1 norm in Equation~\eqref{eqn:onormdef} to see that

\begin{align}
    \norm{\norm{\calL_n}_F^2}_{\psi_1}
    &= \infset{K > 0 : \EE{\exp\left( \norm{\calL_n}_F^2 / K\right) \leq 2 }} \\
    &= \infset{K > 0 : \exp\left( \norm{\opcalP{\bm{A}_n}}_2^2 / K \right) \leq 2 }  \\
    &\leq \infset{K > 0 : \exp\left( \norm{\bm{Q}^T\bm{b}_n}_2^2 + \eta\norm{\bm{V}^T\bm{s}_n}_2^2/K\right) \leq 2 } \\
    &\leq \infset{K > 0 : \exp\left(R\mu_b^2/N + 1/T/K\right) \leq 2} \\
 \end{align}

using the fact that the norm is non-negative, the coherence term and definition of $\bm{s}_n$. It follows that for $K$ large enough that

\begin{equation}
  \norm{\norm{\calL_n}_F^2}_{\psi_1} = \frac{TR\mu_b^2 + N }{NT\log\left(2\right)}  \leq R\frac{T\mu_b^2 + N }{M}
\end{equation}

We now have bounds on the variance and the Orliczs norm and thus can provide a bound on the largest singular value of the operator via Theorem ~\ref{thm:matbern}. Specifically, we can see that the first term in Theorem~\ref{thm:matbern} using $t = \beta\log(TN) > \log(N + T)$ for some $\beta > 0$ and with $C$ large enough

\begin{equation}
    \sigma_X\sqrt{t + \log\left( N + T\right)} \leq \sqrt{2\beta R\frac{\left(N+T \mu_b^2\right)}{M\kappa}\log\left(NT\right)}.
\end{equation}

Similarly, we can bound the second term
\begin{align}
    U_1\log\left(\frac{M U_1^2}{\sigma_X^2}\right)\left(t + \log\left(N + T\right)\right) & \leq 2\beta U_1\log\left(\kappa M\right)\log\left(NT\right)\\
    & \leq 2\beta R\frac{\left(N+T \mu_b^2\right)}{M}\log\left(cR(N + T\mu_b^2)\right)\log\left(NT\right) \\
    & \leq 2\beta R\frac{\left(N+T \mu_b^2\right)}{M}\log^2\left(NT\right).
\end{align}

Putting these two terms together in the Bernstein's inequality yields
\begin{eqnarray}
    \norm{\kappa \mathcal{P}_T\mathcal{A}^T_k\mathcal{A}_k\mathcal{P}_T - \mathcal{P}_T} \leq c\max\left[ R\frac{\left(N+T \mu_b^2\right)}{M}\log^2\left(NT\right), \sqrt{2\beta R\frac{\left(N+T \mu_b^2\right)}{M\kappa}\log\left(NT\right)} \right].
\end{eqnarray}

Now, by assuming that
\begin{equation}
    M \geq C\frac{R}{M}\left(\mu_b^2T + N\right)\log^2(NT),
\end{equation}
and taking the union bound over the ${\kappa}$ partitions, we  complete the proof.

\end{proof}

\subsection{Bounding Certificates}
\begin{lemma}\label{lem:lemma2}
    Let $\calA_k$ be defined as in Equation~\eqref{eqn:linopdef}, $\kappa < M$ be the number of steps in the golfing scheme and assume that $M \leq TN$. Then as long as
    \begin{align}
        M \geq C \kappa R \max\{\mu_bT, N\}\log(NT)
    \end{align}

    then with probability at least $1 - O(M(TN)^{-\beta})$, we have
    \begin{align}
        \max_k\norm{\kappa\calA^T_k\calA_k(\widetilde{\bm{Z}}_{k-1}) - \widetilde{\bm{Z}}_{k-1}} \leq 2^{-(k+1)}. \nonumber
    \end{align}
\end{lemma}

\begin{proof}

The proof is based on bounding the operator norm of $\kappa\mathcal{A}^{T}\mathcal{A} - \mathcal{I}$.
As with the previous lemma, the core of this proof is the use of the matrix Bernstein Inequality~\ref{thm:matbern}. Consider an arbitrary matrix $\bm{G}$ such that,

\begin{equation}
    X_n = \kappa (\dotprod{\bm{G}}{\bm{A}_n}\bm{A}_n - \EE{\dotprod{\bm{G}}{\bm{A}_n}\bm{A}_n}).
\end{equation}

Our goal is to first control $\norm{\sum \EE{X_nX_n^T}}$ and $\norm{\sum \EE{X_n^TX_n}}$. We start with the latter of these terms, $\norm{\sum \EE{X_n^TX_n}}$:
\begin{align}
    \norm{\sum_{n\in\Gamma_k} \EE{X_n^TX_n}}
    &\leq \kappa^2\norm{\sum_{n\in\Gamma_k} \EE{|\dotprod{\bm{G}}{\bm{A}_n}|^2\bm{A}_n\bm{A}_n^*}} \nonumber \\
    &\leq \kappa^2\norm{\sum_{n\in\Gamma_k} \EE{\|\bm{G}\bm{b}\|_2^2\|\bm{s}_n^T\|_2^2\bm{b}_n\bm{s}_n^T\bm{s}_n\bm{b}_n^T}} \nonumber \\
    &\leq \frac{\kappa^2}{T}\max\|\bm{G}\bm{b}\|_2^2\norm{\sum_{n\in\Gamma_k} \EE{\bm{b}_n\bm{b}_n^T}} \nonumber \\
    & = \frac{\mu_bT\kappa^2}{L'^2T^2}\|\bm{G}\|_F^2\norm{\sum_{n\in\Gamma_k} \bm{I}} \nonumber \\
    &\leq \frac{\mu_bT\kappa}{M} \norm{\bm{G}}_F^2
\end{align}

For the first term, $\norm{\sum \EE{X_nX_n^T}}$, we can similarly bound
\begin{align}
    \norm{\sum_{n\in\Gamma_k} \EE{X_nX_n^T}}
    &\leq \kappa^2\norm{\sum_{n\in\Gamma_k} \EE{|\dotprod{\bm{G}}{\bm{A}_n}|^2\bm{A}_n^T\bm{A}_n}} \nonumber \\
    &= \frac{N\kappa^2}{L'^2}\norm{\sum_{n\in\Gamma_k} \EE{\|\bm{G}\bm{s}_n\|_2^2 \bm{s}_n\bm{s}_n^T}} \nonumber \\
    &= \frac{N\kappa^2}{L'^2T^2}\norm{\sum_{n\in\Gamma_k} \EE{\|\bm{G}\bm{s}_n\|_2^2}} \nonumber \\
    &\leq \frac{N\kappa^2}{L'^2T^2} \|\bm{G}\|_F^2\norm{\sum_{n\in\Gamma_k} \bm{I}} \nonumber \\
    &\leq \frac{N\kappa}{M}\norm{\bm{G}}_F^2
\end{align}

Thus, we can write that
\begin{equation}
	\sigma_X^2 \leq \frac{\kappa}{M} \norm{\bm{G}}_F^2\max\{\mu_b^2T, N\}
\end{equation}

To use the Bernstein's inequality, as with Lemma~\ref{lem:lemma1}, we need to bound the Orlicz norm for $\alpha = 1$.
\begin{eqnarray}
    U_1 & = & \norm{X}_{\psi_1} \\
    & \leq & 2\kappa \norm{~\norm{\dotprod{\bm{G}}{\bm{A}_n}\bm{A}_n}_F^2 }_{\psi_1}\\
    & \leq & c\kappa \norm{\dotprod{\bm{G}}{\bm{A}_n}}_{\psi_2}\norm{~\norm{\bm{A}_n}_F }_{\psi_2}  \\
    & \leq & c\kappa \norm{\Tr{\bm{s}_n\bm{b}_n^T\bm{G}}}_{\psi_2}\sqrt{~\norm{~\norm{\bm{b}_n}_2^2\norm{\bm{s}_n}_2^2 }_{\psi_2}} \\
    & \leq & c\kappa \norm{\bm{b}_n^T\bm{G}\bm{s}_n}_{\psi_2} \\
    & \leq & c\kappa \norm{~\norm{\bm{G}^T\bm{b}_n}_2\norm{\bm{s}_n}_2}_{\psi_2}\\
    & \leq & c\kappa \norm{~\norm{\bm{G}^T\bm{b}_n}_2}_{\psi_2}\\
    & \leq & c\kappa\sqrt{\frac{1}{L'}} \norm{\bm{G}}_F \\
    & \leq & c\kappa\sqrt{\frac{T}{M}} \norm{\bm{G}}_F
\end{eqnarray}

We can now apply the matrix Bernstein theorem with the calculated values of $U_1$ and $\sigma_X$. Again, using $t = \log(LN) > \log(N + L)$, the bound is

\begin{eqnarray}
    \sigma_X\sqrt{t + \log(T + N)} &\leq& \kappa \sqrt{\beta \frac{\kappa}{M} \norm{\bm{G}}_F^2\max\{\mu_bT, N\}\log(TN)} \\
    &\leq& \kappa c2^{-k}\sqrt{\beta \frac{\kappa}{M} R\max\{\mu_bT, N\}\log(TN)}.
\end{eqnarray}

where here we use the bound $\norm{\bm{G}}_F\leq 2^{-k}\sqrt{R}$. Similarly,

\begin{eqnarray}
    \ast & = & U_1 \log\left(\frac{\Delta U_1^2}{\sigma^2_X}\right)(t + \log(N + T))  \\
    &  \leq & c\kappa\beta \norm{\bm{G}}_F\sqrt{\frac{T\mu_b^2}{M}}\log \left(c^2\kappa^2\frac{M}{\kappa} \frac{MT \norm{\bm{G}}_F^2 }{M\kappa R\max(\mu_b^2T,N) \norm{\bm{G}}_F^2 } \right)\log(NT)  \\
    &  \leq & c\kappa\beta \norm{\bm{G}}_F\sqrt{\frac{T}{M}}\log \left(c^2 \frac{TM }{\max(\mu_b^2T,N)} \right)\log(NT) \\
    &  \leq & c\kappa\beta 2^{-k}\sqrt{\frac{TR}{M}}\log^2(NT),
\end{eqnarray}
where the last inequality uses the bound $\norm{\bm{G}}_F\leq 2^{-k}\sqrt{R}$. We can now bound $\mu_k^2 \leq \mu_b^2$ with probability $1-O(M(NT)^{-\beta})$ and the prior Lemma to bound $\norm{\bm{G}_k}_F \leq 2^{-k}\sqrt{R}$, which gives us a final bound of

\begin{align}
    \norm{(\calA^*\calA - \mathcal{I})\bm{G}_k}
    & \leq  2^{-k/2} \max\left[  \sqrt{\frac{\kappa\beta R}{M}\max\{\mu_bT, N\}\log(LN)}, c\kappa\beta \sqrt{\frac{RT}{M}}\log^2(NT)\right].
\end{align}

Taking
\begin{equation}
        M\geq C \kappa R \max\{\mu_bT, N\}\log(NT)
\end{equation}
for $C$ large enough and $\kappa=1$ completes the proof.

\end{proof}

\end{document}